\documentclass[conference]{IEEEtran}
\IEEEoverridecommandlockouts

\usepackage{xspace}
\newcommand{\method}{PACE\xspace}
\newcommand{\methodfull}{Partial Accent-Control Editing\xspace}

\newcommand{\secs}{\ensuremath{\mathrm{SECS}}\xspace}

\usepackage[T1]{fontenc}
\usepackage[english]{babel}
\usepackage{times}
\usepackage{latexsym}
\usepackage{inconsolata}
\usepackage{microtype}
\usepackage{graphicx}
\usepackage{booktabs}
\usepackage{multirow}
\usepackage{amsmath,amssymb,amsfonts}
\usepackage{xcolor}
\usepackage{hyperref}

\def\BibTeX{{\rm B\kern-.05em{\sc i\kern-.025em b}\kern-.08em
    T\kern-.1667em\lower.7ex\hbox{E}\kern-.125emX}}
\begin{document}

\title{Partial Accent-Control Editing in Frozen Speech Representations for Accent Conversion}

\author{
\IEEEauthorblockN{
Yangyang Qu,
Michele Panariello,
Massimiliano Todisco,
Nicholas Evans
}
\IEEEauthorblockA{
\textit{EURECOM} \\
Sophia Antipolis, France \\
}
}

\maketitle
\begin{abstract}
Accent conversion is the task of modifying a speech recording so that it sounds closer to a target accent while preserving linguistic content and other speaker-related characteristics. Most accent conversion systems use trained, generative models. Although they can induce target-accented speech, the strength of accent modification is not controllable at inference time, making it difficult to analyse how the strength of accent conversion affects source preservation. We propose \methodfull{} (\method{}), an accent conversion framework based upon the editing of frozen WavLM representations without the training of an accent-conditioned generator. A constrained edit is first applied to source WavLM features, which are then fused with target-accent reference features retrieved from non-parallel accent examples. Fusion weights are used to control the trade-off between accent conversion strength and the degradation of other source attributes. 
Using a suite of five metrics, and with the cost of accent-entangled speaker similarity, we show that \method{} is substantially superior to a pair of competitive baselines in terms of both accent conversion and source preservation. 
\end{abstract}


\section{Introduction}

Accent conversion aims to modify a source utterance so that it sounds closer to a target accent while preserving its linguistic content and speaker-related characteristics. This paper studies this task in a non-parallel setting. Target-accent reference utterances are available, but they do not share the same transcript as the source utterance. The goal is to use these non-parallel examples to guide the accent change without relying on a parallel target recording.

Recent work on accent conversion and accent-aware speech generation has mainly relied on trained generative models. Representative directions include synthetic-pair or text-to-speech(TTS)-guided accent conversion, disentanglement-based modeling, and accent-aware multispeaker TTS~\cite{nguyen2022accent, nguyen2025improving, nguyen2025streaming, huang2024usd, melechovsky2024dart, melechovsky2024accent}. These systems can produce target-accented speech, but the strength of accent modification is typically determined by the trained generator, latent representation, or conditioning input. As a result, accent-modification strength is not usually exposed as a direct inference-time control.

This lack of direct control makes it difficult to study accent conversion as an operating-point trade-off. Stronger target-accent cues may improve target-accent recognition, but they may also affect transcript consistency, intelligibility, and speaker similarity. Existing systems can be evaluated with preservation metrics, but they do not always provide a simple way to vary target-accent strength within the same conversion mechanism. A useful controlled conversion framework should therefore allow target-accent strength to be adjusted and its effect on source preservation to be measured.

Motivated by this goal, we propose \methodfull{} (\method), a generator-free framework for controllable accent conversion in frozen self-supervised speech representations. Instead of training a new accent-conditioned conversion generator, \method edits WavLM~\cite{chen2022wavlm} representations extracted from a frozen encoder. It first applies a constrained source-side edit in an accent-predictive subspace. It then performs retrieval by selecting nearest-neighbour target-accent frames from a non-parallel reference bank and fuses the corresponding reference features with the edited source features. The fusion weight controls the balance between the edited source features and the retrieved target-accent features, providing explicit inference-time control over target-accent influence.

Experiments with the L2-ARCTIC~\cite{zhao2018l2} database show that this control exposes a clear trade-off between classifier-based target-accent accuracy and source preservation. Increasing the influence of retrieved target-accent features improves target-accent accuracy measured by a pretrained classifier, but also increases preservation cost, especially in intelligibility and speaker similarity. To understand the source of this trade-off, we conduct ablations on the transport-based edit, retrieval space, and fusion weight. We also compare \method with reproduced accent-aware generative references under a shared evaluation protocol. These results support the view that non-parallel accent conversion can be studied as controllable partial editing in frozen speech representations.

Our main contributions are as follows:
\begin{itemize}
    \item We formulate non-parallel accent conversion as controllable partial editing in frozen self-supervised speech representations, where target-accent influence can be adjusted at inference time rather than fixed by a trained generator.
    \item We propose \method{}, a generator-free framework that combines an accent-predictive source update with retrieval-based target-accent fusion in frozen WavLM space.
    \item We evaluate \method{}  on L2-ARCTIC, showing how fusion strength and retrieval space affect the trade-off between classifier-based target-accent evidence and source preservation.
\end{itemize}

\begin{figure*}[t]
    \centering
    \includegraphics[width=\linewidth]{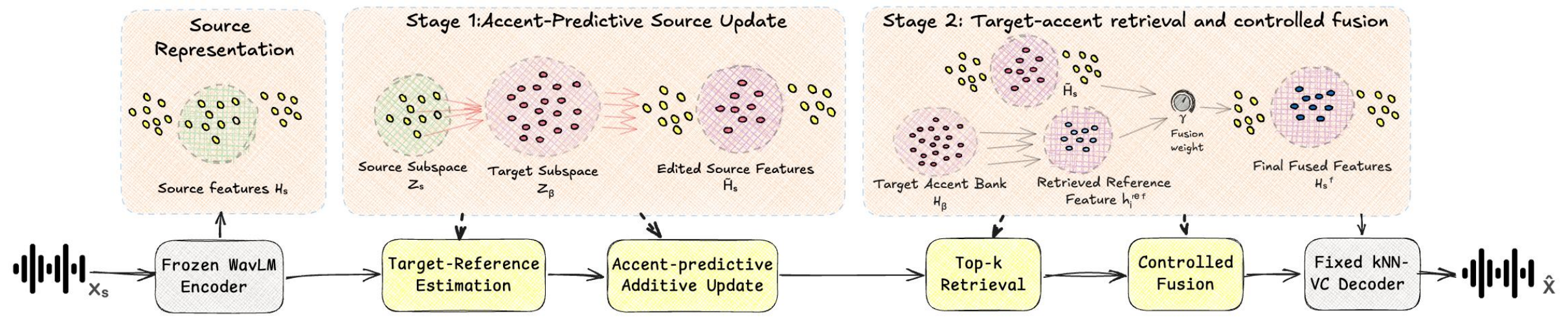}
    \caption{Overview of \method{}. Stage~1 performs an accent-predictive source update by estimating a transport-based target reference in the accent-predictive subspace and applying the resulting shift as an additive update to the source WavLM features. Stage~2 performs top-$k$ retrieval in the accent-predictive subspace, uses the retrieved indices to fetch WavLM-space target-accent reference features from the bank, and fuses them with the edited source features using the fusion weight $\gamma$. The final representation is synthesized with the fixed kNN-VC synthesis backend.
}
    \label{fig:architecture}
\end{figure*}

\section{Related Work}

Prior work on accent conversion and accent-aware speech generation differs mainly in how accent information is represented and introduced into speech. We review four directions: phonetic, articulatory, and token-based representations; disentangled representations; synthetic-pair or text-to-speech-guided supervision; and accent-aware generative or retrieval-based synthesis.

\subsection{Phonetic, Articulatory, and Token-Based Representations}

Early accent-conversion methods based on phonetic posteriorgrams (PPGs) treat accent conversion as pronunciation-level modification in an explicit phonetic representation~\cite{zhao2019foreign}. Recent work also uses articulatory representations to capture accent-related variation~\cite{siriwardena2024accent}, or intermediate speech units such as discrete units synthesized from controllable accented text-to-speech systems~\cite{nguyen2024discrete} and self-supervised discrete tokens with non-parallel data~\cite{bai2025accent}. These representations are useful for accent modeling, but may still encode speaker characteristics, prosody, or acoustic realization. In contrast, \method{} edits frozen WavLM representations and restricts the initial source update to accent-predictive directions estimated from labeled data.

\subsection{Disentangled Representations}

Another line of work uses disentangled or factorized speech representations to separate accent from content, speaker identity, or prosody~\cite{quamer2023decoupling,huang2024usd}. Such separation is attractive for accent conversion, but strong disentanglement is difficult to guarantee because accent, speaker traits, content, and prosody are often correlated in learned representations. \method{} does not assume full disentanglement; instead, it studies accent conversion as a controlled trade-off between classifier-based target-accent evidence and source preservation.

\subsection{Synthetic-Pair and Text-to-Speech-Guided Accent Conversion}

Because parallel accent-conversion data are scarce, recent systems often use synthetic pairs, auxiliary text-to-speech systems, or teacher-guided training~\cite{nguyen2022accent,nguyen2025improving,nguyen2025streaming}. These methods make accent conversion more supervised, but can depend on the quality of synthetic targets, teacher models, and alignment pipelines. \method{} addresses a different setting: it uses accent-labeled data and non-parallel target-accent examples, without requiring synthetic parallel target utterances or training a new accent-conditioned generator.

\subsection{Accent-Aware Generative and Retrieval-Based Synthesis}

Accent has also been studied in multi-accent text-to-speech and generative synthesis. DART models accent and speaker factors in multispeaker text-to-speech~\cite{melechovsky2024dart}, while MLV performs accent conversion in text-to-speech using a multi-level VAE and adversarial training~\cite{melechovsky2024accent}. Related work further studies transliteration-based multi-accent synthesis~\cite{inoue2025macst}, semantic-token conversion followed by target-accent generation~\cite{jia2024convert}, and parameter-efficient accent adaptation in text-to-speech~\cite{yang2023parameter}.

Retrieval-based voice conversion is also related to \method{}. kNN-VC replaces source frames with nearest-neighbour reference frames in self-supervised representation space and synthesizes the result with a pretrained vocoder~\cite{baas2023knnvc}. \method{} adopts a related retrieval view, but retrieves from a target-accent bank rather than a target-speaker reference, performs retrieval in an accent-predictive subspace, and uses a fusion weight to control target-accent influence. Overall, \method{} differs from prior work by editing frozen WavLM representations and exposing an inference-time trade-off between classifier-based target-accent evidence and source preservation.

\section{Proposed Method}
\label{sec:architecture}

\subsection{Overview and problem setup}

Figure~\ref{fig:architecture} summarizes the proposed \method{} framework. Given a source waveform $x_s$ and a target accent label $a_t$, the goal is to generate a converted waveform $\hat{x}$ that sounds closer to the target accent while preserving the linguistic content and speaker-related characteristics of $x_s$.

We study the non-parallel setting, where target-accent utterances are available but do not share transcripts with the source utterance. \method{} uses these utterances to build a frame-level target-accent reference bank in frozen WavLM space, avoiding the need for paired target waveforms.

\method{} performs accent conversion through two representation-level stages. Stage~1 applies a source-side update in an accent-predictive subspace. Stage~2 retrieves target-accent reference features from the bank and fuses them with the edited source features using a controllable fusion weight. The final edited WavLM sequence is then converted back to waveform with a fixed kNN-VC~\cite{baas2023knnvc} synthesis backend.

\subsection{Frozen WavLM representations}

We represent the source utterance $x_s$ as a sequence of frozen WavLM frame embeddings. The frozen WavLM encoder $f(\cdot)$ maps $x_s$ to
\begin{equation}
H_s = f(x_s) = [h_1^s,\dots,h_n^s]^\top \in \mathbb{R}^{n \times D},
\end{equation}
where $n$ is the number of WavLM frames, $D$ is the WavLM feature dimension, and $h_i^s \in \mathbb{R}^{D}$ is the source frame embedding at frame $i$.

Frozen WavLM representations contain phonetic, speaker-related, and accent-related cues in a shared feature space. To avoid applying an unrestricted update in the full WavLM space, \method{} first defines an accent-predictive subspace. We estimate a projection matrix $P_{\mathrm{ac}}$ once from the accent-labeled training split and keep it fixed during conversion. After standardizing WavLM features with training-set statistics, we apply partial least squares (PLS) between frame-level WavLM features and one-hot accent labels. The resulting directions are predictive of accent labels. We orthonormalize these directions and use them as the rows of
$
P_{\mathrm{ac}} \in \mathbb{R}^{d_{\mathrm{ac}} \times D},
$
where $d_{\mathrm{ac}}$ is the dimension of the accent-predictive subspace.

For a WavLM frame embedding $h \in \mathbb{R}^{D}$, its projection into this subspace is
$
z = P_{\mathrm{ac}} h \in \mathbb{R}^{d_{\mathrm{ac}}}.
$
We refer to this space as accent-predictive because its directions are estimated from accent labels. This term does not imply that the space contains only accent information.

\subsection{Target-accent reference bank}
\label{subsec:target-bank}

For the requested target accent label $a_t$, \method{} builds a frame-level reference bank from training utterances with that accent. The bank contains WavLM frame embeddings extracted from multiple target-accent utterances. It is used because the available target-accent utterances are non-parallel to the source utterance and therefore cannot provide direct frame-by-frame targets.

We denote the target-accent bank as
$
H_b = [h_1^b,\dots,h_m^b]^\top \in \mathbb{R}^{m \times D},
$
where the superscript $b$ denotes bank frames and $m$ is the number of frames in the bank for accent $a_t$. Each row $h_j^b \in \mathbb{R}^{D}$ is a WavLM frame embedding from a target-accent training utterance.

The same bank serves two roles. In Stage~1, it provides target-accent frames for estimating an initial shift in the accent-predictive subspace. In Stage~2, its projected features are used for top-$k$ retrieval in the accent-predictive subspace, while the corresponding WavLM-space bank features are averaged to form reference features and then fused with the edited source features.

\subsection{Stage~1: Accent-predictive source update}
\label{subsec:stage1}

Stage~1 gives the source representation an initial shift toward the target accent. The update is estimated in an accent-predictive subspace and then mapped back to the WavLM feature space, so that the source frames are modified only along directions selected from accent-labeled data.

We first project the source sequence $H_s$ and the target-accent bank $H_b$ into the accent-predictive subspace:
\begin{equation}
\label{eq:accent-subspace-projection}
    Z_s = H_s P_{\mathrm{ac}}^\top \in \mathbb{R}^{n \times d_{\mathrm{ac}}},
    \qquad
    Z_b = H_b P_{\mathrm{ac}}^\top \in \mathbb{R}^{m \times d_{\mathrm{ac}}}.
\end{equation}
Here, the $i$-th row of $Z_s$ and the $j$-th row of $Z_b$ denote the projected source frame $z_i^s$ and target-accent bank frame $z_j^b$, respectively.

\subsubsection{Transport-based target-reference estimation}

Because the source utterance and the target-accent bank are non-parallel, there is no frame-level alignment between their WavLM representations. We therefore use entropy-regularized unbalanced optimal transport~\cite{cuturi2013sinkhorn,chizat2018unbalanced} to estimate a weighted matching between the projected source sequence $Z_s$ and the projected target-accent bank $Z_b$.

Let
$
    \Pi \in \mathbb{R}_{\ge 0}^{n \times m}
$
denote the transport plan, where $\Pi_{ij}$ is the matching weight between source frame $z_i^s$ and target-accent bank frame $z_j^b$. A larger value of $\Pi_{ij}$ means that bank frame $j$ contributes more to the target-accent reference for source frame $i$. We define the frame-level matching cost as
$
    C_{ij} = \lVert z_i^s - z_j^b \rVert_2^2 .
$

The transport plan is obtained by minimizing
\begin{equation}
\label{eq:uot-objective}
\begin{aligned}
\mathcal{L}(\Pi)
&= \langle C, \Pi \rangle  + \lambda \Bigl(
\mathrm{KL}\bigl(r(\Pi)\parallel \mathbf{w}_s\bigr)
+
\mathrm{KL}\bigl(c(\Pi)\parallel \mathbf{w}_b\bigr)
\Bigr) \\
&\quad + \varepsilon\,
\mathrm{KL}\bigl(\Pi \parallel \mathbf{w}_s \mathbf{w}_b^\top\bigr),
\end{aligned}
\end{equation}
where $\lambda$ controls the penalty on deviations from the initial frame weights, and $\varepsilon$ controls the entropy regularization. The vectors $\mathbf{w}_s$ and $\mathbf{w}_b$ assign equal initial mass to source frames and bank frames, respectively; each source frame has weight $1/n$, and each bank frame has weight $1/m$. The row marginal $r(\Pi)$ collects the total matching weight assigned to each source frame, while the column marginal $c(\Pi)$ collects the total matching weight received by each bank frame.

The first term in Eq.~\eqref{eq:uot-objective} favors matching source and bank frames that are close in the accent-predictive subspace. The unbalanced KL terms penalize large deviations from the initial frame weights, but they do not require exact mass conservation. This is useful for non-parallel data, where some source frames may not have close counterparts in the target-accent bank and some bank frames may not be used. The entropy term smooths the transport plan and improves optimization stability. In practice, we minimize Eq.~\eqref{eq:uot-objective} with the unbalanced Sinkhorn solver implemented in POT~\cite{flamary2021pot}.
 
\subsubsection{Accent-predictive additive update}
The transport plan defines, for each source frame, a transport-weighted target-accent reference in the WavLM feature space. Let $\Pi^\star$ be the optimized transport plan. For source frame $i$, we compute
\begin{equation}
\label{eq:weighted-bank-feature}
    \bar{h}_i^b
    =
    \frac{\sum_{j=1}^{m} \Pi^\star_{ij} h_j^b}
    {\sum_{j=1}^{m} \Pi^\star_{ij} + \delta},
\end{equation}
where $\delta$ is a small constant used for numerical stability. The vector $\bar{h}_i^b$ is a weighted average of target-accent bank frames matched to source frame $i$.

We then compute the target-directed shift in the accent-predictive subspace:
\begin{equation}
\Delta z_i = P_{\mathrm{ac}}\bar{h}_i^b - z_i^s .
\end{equation}
This vector indicates how source frame $i$ shifts toward its transport-weighted target-accent reference after both are projected into the accent-predictive subspace.

Finally, we map this shift back to the WavLM feature dimension and add it to the original source frame:
$
\label{eq:additive-update}
    \tilde{h}_i
    =
    h_i^s + \alpha P_{\mathrm{ac}}^\top \Delta z_i ,
$
where $\alpha \in [0,1]$ controls the scale of the Stage~1 update. Because the rows of $P_{\mathrm{ac}}$ are orthonormalized, the update term $P_{\mathrm{ac}}^\top \Delta z_i$ changes the source frame only along the selected accent-predictive directions.

Stacking the updated frames gives the Stage~1 edited source sequence:
\begin{equation}
\tilde{H}_s =
[\tilde{h}_1,\dots,\tilde{h}_n]^\top
\in \mathbb{R}^{n \times D}.
\end{equation}
This sequence is the input to Stage~2.
\subsection{Stage~2: Target-accent retrieval and controlled fusion}
\label{subsec:stage2}

Stage~2 controls how much target-accent reference information is introduced after the Stage~1 source update. The retrieval step is performed in the accent-predictive subspace, while the retrieved reference features are fetched from the original WavLM-space bank and fused with the edited source features.

We first project the Stage~1 edited source sequence into the accent-predictive subspace:
\begin{equation}
    \tilde{Z}_s = \tilde{H}_s P_{\mathrm{ac}}^\top \in \mathbb{R}^{n \times d_{\mathrm{ac}}}.
\end{equation}
The projected target-accent bank $Z_b$ is the same bank projection defined in Eq.~\eqref{eq:accent-subspace-projection}. For each edited source frame $\tilde{z}_i^s$, we search the projected target-accent bank $Z_b$ and select its top-$k$ nearest bank frames under cosine similarity:
\begin{equation}
    \mathcal{N}_k(i) =
    \operatorname{TopK}_{j \in \{1,\ldots,m\}}
    \operatorname{cos}(\tilde{z}_i^s, z_j^b).
\end{equation}
This search returns bank indices rather than final reference features. We then use these indices to gather the corresponding WavLM-space frames from the original target-accent bank $H_b$:
\begin{equation}
    h_i^{\mathrm{ref}}
    =
    \frac{1}{k}
    \sum_{j \in \mathcal{N}_k(i)}
    h_j^b
    \in \mathbb{R}^{D}.
\end{equation}
Thus, the accent-predictive subspace is used to determine which target-accent frames to retrieve, whereas the retrieved reference feature remains in the original WavLM feature space.

We then fuse the edited source frame and the retrieved reference feature:
\begin{equation}
    h_i^f = (1-\gamma)\tilde{h}_i + \gamma h_i^{\mathrm{ref}},
\end{equation}
where $\gamma \in [0,1]$ is the fusion weight. When $\gamma=0$, the final frame remains the Stage~1 edited source frame. When $\gamma=1$, the final frame is replaced by the retrieved target-accent reference. Intermediate values interpolate between these two representations, controlling the balance between source preservation and target-accent influence.

Stacking the fused frames gives the final edited WavLM sequence:
\begin{equation}
    H_s^f = [h_1^f,\ldots,h_n^f]^\top \in \mathbb{R}^{n \times D}.
\end{equation}
The final edited WavLM sequence is synthesized with the fixed kNN-VC synthesis backend, which consists of a mel projection and a HiFi-GAN vocoder.  These synthesis modules are kept frozen, so \method{} performs accent conversion only by editing the WavLM representation rather than by training a new accent-conditioned waveform generator.

\begin{table*}[t]
\centering
\small
\setlength{\tabcolsep}{7pt}
\caption{
Main operating points on the full 5k protocol. Accuracy is computed with a pretrained accent classifier and is reported as a percentage. The rows are ordered from more source-preserving to more target-accent-oriented settings.
}
\label{tab:main_operating_points}
\begin{tabular}{lccccc}
\toprule
Operating point & WER(\%)$\downarrow$ & CER(\%)$\downarrow$ & STOI$\uparrow$ & \secs{}$\uparrow$ & Accuracy(\%)$\uparrow$ \\
\midrule
Stage~1 only                & 14.2 &  7.4 & 0.884 & 0.833 &  4.4 \\
Controlled fusion (default) & 16.0 &  8.3 & 0.802 & 0.583 & 39.0 \\
Stage~2 only                & 16.2 &  8.5 & 0.799 & 0.584 & 38.7 \\
Original WavLM retrieval    & 18.2 &  9.5 & 0.772 & 0.470 & 52.7 \\
Reference replacement       & 20.1 & 10.4 & 0.750 & 0.413 & 66.5 \\
\bottomrule
\end{tabular}
\end{table*}
\section{Experiments}
\label{sec:experiments}

\subsection{Dataset and evaluation protocol}
\label{subsec:data}

We evaluate \method{} on L2-ARCTIC~\cite{zhao2018l2}, a non-native English speech corpus with speaker-level L1-background annotations. The corpus includes six L1-background accent labels: Arabic, Hindi, Korean, Mandarin, Spanish, and Vietnamese. We use these labels as the target accents in our evaluation.

Our main protocol contains 5,000 source-target conversion pairs. We select 1,000 source utterances and convert each one to the five target accents different from its own L1-background accent, yielding a balanced cross-accent evaluation without same-accent conversions. For each target accent, the reference bank is built only from utterances spoken by speakers with that target accent; therefore, the source utterance and its target-accent bank are speaker-disjoint for every conversion pair.

\subsection{Implementation details}
\label{subsec:implementation}

We use WavLM-Large as the frozen SSL encoder and extract 1024-dimensional frame-level features from layer~6. The edited WavLM sequence is synthesized with the fixed kNN-VC feature-to-waveform backend, consisting of a mel projection and a HiFi-GAN vocoder~\cite{baas2023knnvc}. Both synthesis modules remain frozen in all experiments. All waveforms are synthesized and evaluated at 16~kHz.

For Stage~1, we use entropy-regularized unbalanced optimal transport with $\lambda=0.1$ and $\epsilon=0.01$. For Stage~2, each target-accent bank contains 50k frames after subsampling. Unless otherwise stated, top-$k$ retrieval is performed in the accent-predictive subspace with $k=4$, and the
default \method{} setting uses $\alpha=1.0$ and $\gamma=0.7$.

\subsection{Operating points and reproduced references}
\label{subsec:operating-points}

We evaluate the following operating points.

\textbf{Stage~1 only} sets $\gamma=0$, so the final representation is the Stage~1 edited source sequence without target-accent fusion. This setting measures the effect of the accent-predictive source update alone.

\textbf{Controlled fusion} uses the default setting $\alpha=1.0$, top-$k=4$, and $\gamma=0.7$. Retrieval is performed in the accent-predictive subspace, and the retrieved indices are used to fetch WavLM-space target-accent reference features from $H_b$. These reference features are then fused with the Stage~1 edited source features. This is the default \method{} operating point.

\textbf{Stage~2 only} sets $\alpha=0$ while keeping the same retrieval and fusion procedure. This setting removes the Stage~1 source update and tests how much of the final performance comes from Stage~2 retrieval and fusion alone.

\textbf{Original WavLM retrieval} performs Stage~2 nearest-neighbour retrieval directly in the original WavLM feature space, instead of using the accent-predictive subspace for retrieval. The remaining reference fetching and fusion steps are kept unchanged. This setting tests whether full-space retrieval increases target-accent evidence and how it affects source preservation.

\textbf{Reference replacement} sets $\gamma=1$, so each final frame is replaced by the retrieved target-accent reference feature. This setting represents the strongest target-accent reference influence among the reported operating points.

For shared-protocol comparison, we reproduce two accent-aware generative references, DART~\cite{melechovsky2024dart} and MLV~\cite{melechovsky2024accent}. Direct comparison is limited in much of the recent accent-conversion literature, where systems are often evaluated under method-specific datasets, training conditions, or metrics. To provide a more controlled comparison, we train these two reference systems following their publicly released implementations and evaluate them using the same dataset, conversion protocol, and metrics as \method{}.
\subsection{Evaluation metrics}
\label{subsec:metrics}

We evaluate converted speech from two perspectives: source preservation and classifier-based target-accent evidence. For source preservation, we report word error rate (WER) and character error rate (CER) between Whisper-base transcripts of the source and converted speech~\cite{radford2023robust}. Lower WER and CER indicate better transcript consistency.

We also report short-time objective intelligibility (STOI)~\cite{taal2011algorithm} and speaker embedding cosine similarity (\secs) computed with Resemblyzer~\cite{wan2018generalized}. Higher STOI and \secs indicate stronger preservation of intelligibility and speaker-related characteristics.

For target-accent evaluation, we use a pretrained accent classifier.\footnote{\url{https://huggingface.co/kaysrubio/accent-id-distilhubert-finetuned-l2-arctic2}} For each converted utterance, the classifier predicts one accent label. We define target-accent accuracy as the percentage of converted utterances whose predicted label matches the intended target accent. We use it to measure whether the converted speech contains accent cues that are detectable by the classifier. Throughout the results, classifier-based target-accent evidence refers to this classifier
output rather than to a human perceptual judgement of accent strength.

\section{Results}
\label{sec:results}

We present four sets of results: main operating points, the effect of the fusion weight, target-accent-specific accuracy, and comparison with reproduced accent-aware references.

\subsection{Main operating points}
\label{subsec:main-results}

Table~\ref{tab:main_operating_points} shows a trade-off between classifier-based target-accent accuracy and source preservation. Stage~1 only gives the strongest preservation, with 14.2\% WER, 0.884 STOI, and 0.833 \secs{}, but achieves only 4.4\% accuracy. This suggests that the Stage~1 accent-predictive source update is conservative and introduces limited target-accent evidence by itself.

The default controlled-fusion setting increases accuracy to 39.0\%, at the cost of lower preservation. Stage~2 only gives a very similar operating point, with 38.7\% accuracy, suggesting that retrieval-based fusion accounts for most of the default setting at $\gamma=0.7$. This suggests that, at $\gamma=0.7$, retrieval-based fusion accounts for most of the aggregate target-accent effect, while the Stage~1 source update mainly provides a conservative source-side initialization. Original WavLM retrieval further improves accuracy to 52.7\%, but reduces \secs{} to 0.470, indicating that full-space retrieval introduces stronger target-accent evidence together with greater preservation loss. Reference replacement reaches the highest accuracy, 66.5\%, but gives the weakest preservation. Overall, \method{} is best viewed as an operating-point framework for balancing target-accent evidence and source preservation.

\subsection{Effect of the fusion weight \texorpdfstring{$\gamma$}{gamma}}
\label{subsec:gamma-results}

Table~\ref{tab:gamma_effect_full} confirms that $\gamma$ provides direct control over the conversion operating point. In controlled fusion, increasing $\gamma$ from 0.3 to 0.7 raises accuracy from 10.6\% to 39.0\%, while WER increases from 12.9\% to 16.0\% and \secs{} decreases from 0.770 to 0.583. The same trend appears in Stage~2 only, showing that $\gamma$ mainly controls the influence of retrieved target-accent reference features.

Original WavLM retrieval follows the same trend but shifts the system toward stronger target-accent evidence. At $\gamma=0.7$, it reaches 52.7\% accuracy, compared with 39.0\% for controlled fusion, but gives weaker preservation. Thus, both the fusion weight and the retrieval space determine the trade-off between classifier-based target-accent evidence and source preservation.

\begin{table}[t]
\centering
\small
\setlength{\tabcolsep}{4.5pt}
\caption{
Effect of the Stage~2 fusion weight $\gamma$ on the full 5k protocol with top-$k=4$. Larger $\gamma$ gives more weight to the retrieved target-accent reference feature. Accuracy is classifier-based and is reported as a percentage.
}
\label{tab:gamma_effect_full}
\begin{tabular}{lccccc}
\toprule
Setting & WER(\%)$\downarrow$ & CER(\%)$\downarrow$ & STOI$\uparrow$ & \secs{}$\uparrow$ & Accuracy(\%)$\uparrow$ \\
\midrule
\multicolumn{6}{l}{\textit{Controlled fusion}} \\
$\gamma=0.3$ & 12.9 & 6.7 & 0.849 & 0.770 & 10.6 \\
$\gamma=0.5$ & 14.1 & 7.2 & 0.832 & 0.697 & 20.6 \\
$\gamma=0.7$ & 16.0 & 8.3 & 0.802 & 0.583 & 39.0 \\
\midrule
\multicolumn{6}{l}{\textit{Stage~2 only}} \\
$\gamma=0.3$ & 13.0 & 6.6 & 0.849 & 0.769 & 10.8 \\
$\gamma=0.5$ & 13.9 & 7.1 & 0.832 & 0.698 & 20.5 \\
$\gamma=0.7$ & 16.2 & 8.5 & 0.799 & 0.584 & 38.7 \\
\midrule
\multicolumn{6}{l}{\textit{Original WavLM retrieval}} \\
$\gamma=0.3$ & 15.2 & 8.0 & 0.821 & 0.617 & 23.5 \\
$\gamma=0.5$ & 16.5 & 8.6 & 0.796 & 0.544 & 38.8 \\
$\gamma=0.7$ & 18.2 & 9.5 & 0.772 & 0.470 & 52.7 \\
\bottomrule
\end{tabular}
\end{table}

\subsection{Target-accent-specific accuracy}
\label{subsec:target-results}

\begin{figure}[t]
\centering
\includegraphics[width=0.9\linewidth]{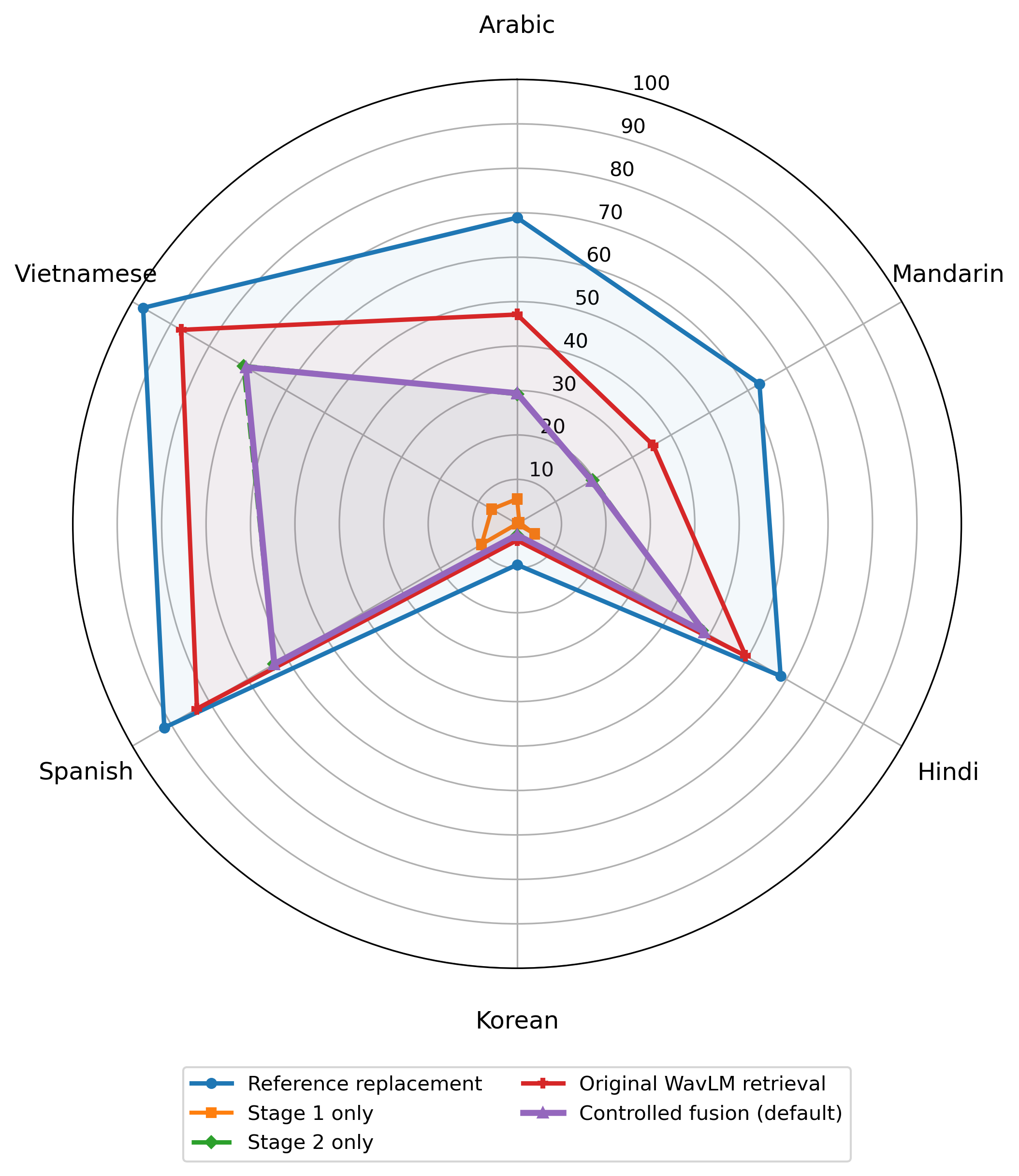}
\caption{
Target-accent-specific classifier-based accuracy for the operating points in Table~\ref{tab:main_operating_points}. Accuracy is reported as a percentage for each intended target accent.
}
\label{fig:target_acc_radar}
\end{figure}

Figure~\ref{fig:target_acc_radar} shows that the aggregate accuracy in Table~\ref{tab:main_operating_points} hides substantial target-accent variation. Spanish and Vietnamese benefit strongly from stronger target-accent reference influence, whereas Korean remains difficult across all operating points. Even under reference replacement, Korean reaches only 9.2\% accuracy, compared with 91.7\% for Spanish and 97.2\% for Vietnamese. This suggests that target-accent-specific results are necessary for interpreting aggregate conversion accuracy.

\subsection{Comparison with reproduced generative references}
\label{subsec:external-results}

Table~\ref{tab:external_baselines} compares the default controlled-fusion setting of \method{} with reproduced DART~\cite{melechovsky2024dart} and MLV~\cite{melechovsky2024accent} under the same protocol. \method{} obtains lower WER and CER, higher STOI, and higher classifier-based accuracy than both reproduced references. It achieves 16.0\% WER and 39.0\% accuracy, compared to 29.7\% WER and 28.2\% accuracy for DART, and 36.8\% WER and 24.3\% accuracy for MLV.

The cost is reduced speaker similarity: 0.583 \secs{} for \method{} and 0.824 for DART. 
With accent being naturally entangled, a \emph{reduction in speaker similarity} might be interpreted as additional evidence of \emph{stronger accent conversion}. 
The default \method{} operating point still provides stronger transcript consistency, intelligibility, and classifier-based target-accent evidence than the baselines.

\begin{table}[t]
\centering
\footnotesize
\setlength{\tabcolsep}{3pt}
\caption{
Shared-protocol comparison with reproduced accent-aware generative references. Here, \method{} denotes the default controlled-fusion operating point. WER, CER, and Accuracy are reported as percentages.
}
\label{tab:external_baselines}
\begin{tabular}{lccccc}
\toprule
Method & WER(\%)$\downarrow$ & CER(\%)$\downarrow$ & STOI$\uparrow$ & \secs{}$\uparrow$ & Accuracy(\%)$\uparrow$ \\
\midrule
DART~\cite{melechovsky2024dart} & 29.7 & 15.7 & 0.202 & \textbf{0.824} & 28.2 \\
MLV~\cite{melechovsky2024accent} & 36.8 & 20.9 & 0.174 & 0.528 & 24.3 \\
\method{} & \textbf{16.0} & \textbf{8.3} & \textbf{0.802} & 0.583 & \textbf{39.0} \\
\bottomrule
\end{tabular}
\end{table}
\section{Conclusion}

We introduced \method{}, a controllable accent-conversion framework that edits frozen WavLM representations rather than training a new accent-conditioned waveform generator. The main idea is to treat non-parallel accent conversion as partial representation editing, where retrieved target-accent references can be incorporated with an adjustable inference-time strength.

Experiments on L2-ARCTIC show that this control exposes a consistent trade-off between classifier-based target-accent evidence and source preservation. Compared to original data, higherthe fusion weights improve classifier-based accuracy at the cost of transcript consistency, intelligibility, and speaker similarity. Our approach is nonetheless superior to representative baselines in almost all considered respects.  The only exception is for speaker similarity. Being inherently-entangled with the speaker accent, this result may stem only from stronger accent conversion.  The shared-protocol comparison with reproduced generative references further suggests that PACE provides a competitive operating point to balance target-accent evidence and source preservation.

A limitation of the current study is that target-accent evaluation relies on an automatic classifier. Future work will include human listening tests, stronger source-preserving representation edits, and extensions of the partial-editing framework to unseen target accents and other speech attributes.

\section*{Acknowledgement}
This work was supported by the French Agence Nationale de la Recherche (ANR) via the SpeechPrivacy (ANR-23-CE23-0022) and GOOD-BIAS (ANR-25-CE39-6459) projects.
Generative AI tools were used for language editing and stylistic refinement of selected parts of the manuscript.
All scientific claims, experimental results, analyses, and final manuscript content were reviewed and verified by the authors.

\newpage
\bibliographystyle{IEEEtran} 
\bibliography{custom}

\end{document}